# Was Venus the First Habitable World of our Solar System?

**M. J. Way[1,2], Anthony D. Del Genio[1], Nancy Y. Kiang[1], Linda E. Sohl[1,3], David H. Grinspoon[4], Igor Aleinov[1,3], Maxwell Kelley[1], and Thomas Clune[5]**

[1] NASA Goddard Institute for Space Studies, 2880 Broadway, New York, New York, USA.

[2] Department of Astronomy & Space Physics, Uppsala University, Uppsala, Sweden.

[3] Center for Climate Systems Research, Columbia University, New York, New York, USA.

[4] Planetary Science Institute, Tuscon, Arizona, USA.

[5] Global Modeling and Assimilation Office, NASA Goddard Space Flight Center, Greenbelt, Maryland, USA.

Corresponding author: Michael J. Way (michael.j.way@nasa.gov)

**Key Points:**

- Venus may have had a climate with liquid water on its surface for approximately 2 billion years.

- The rotation rate and topography of Venus play crucial roles in its atmospheric dynamics.

- Venus's climatic history has important implications for exoplanetary studies of the habitable zone.

## Abstract

Present-day Venus is an inhospitable place with surface temperatures approaching 750K and an atmosphere over 90 times as thick as present day Earth's. Billions of years ago the picture may have been very different. We have created a suite of 3-dimensional climate simulations using topographic data from the Magellan mission, solar spectral irradiance estimates for 2.9 and 0.715 billion years ago, present-day Venus orbital parameters, an ocean volume consistent with current theory and measurements, and an atmospheric composition estimated for early Venus. Using these parameters we find that such a world could have had moderate temperatures if Venus had a rotation period slower than about 16 Earth days, despite an incident solar flux 46−70% higher than modern Earth receives. At its current rotation period of 243 days, Venus's climate could have remained habitable until at least 715 million years ago if it hosted a shallow primordial ocean. These results demonstrate the vital role that rotation and topography play in understanding the climatic history of exoplanetary Venus-like worlds being discovered in the present epoch.

## 1 Introduction

The habitability states of Earth and Mars have changed through the five billion year history of our solar system [*Sleep* 1994, *Sleep & Zahnle* 2001, *Sleep & Hessler* 2006, *Sleep* 2010, *DiAchille & Hynek* 2010, *Ehlmann* 2011, *Arndt & Nisbet* 2012]. Given the similarity of Venus's size and bulk density, and its proximity to Earth, it is plausible that Venus formed with a similar bulk composition and initial volatile inventory, although other scenarios have been proposed [e.g. *Chassefiere et al.* 2012]. At the same time, Venus has a D/H ratio that is 150±30 times that of terrestrial water [*Donahue et al.* 1982, 1997], but is currently a parched world with atmospheric $H_2O$ of $6x10^{15}$ kg, compared to Earth's surface inventory of $1.4x10^{21}$ kg [*Lecuyer & Guyot* 2000]. The D/H ratio implies that Venus has lost substantial quantities of water over its history, but it is unclear when and at what rate [*Kasting & Pollack* 1983, *Donahue* 1999]. When continuing exogenous and endogenous sources are included, the primordial value is very poorly determined [*Grinspoon* 1993]. However, modern formation models indicate a great deal of mixing among terrestrial planet protoplanets [e.g. *Morbidelli et al.* 2012], strongly suggesting that Venus and Earth did not form with 5 orders of magnitude difference in water inventory.

A variety of estimates give paleo-Venus the equivalent of $4 - 525$ meters of liquid water if spread evenly across its surface [*Donahue & Russell* 1997], while the idea of an ancient Venus with oceans is hardly new [e.g. *Kasting et al.* 1984, *Kasting* 1988, *Grinspoon & Bullock* 2003]. At the same time, some planet formation scenarios for Venus-type objects close to their parent star indicate that most water may have been expelled in the first 100 Myr of their history [*Hamano et al.* 2013]. Recent work demonstrates that Earth may have also lost much of its initial volatile inventory within the first 100 Myr of its history [*Finlay et al.* 2016]. For Venus to obtain an ocean depth of 100s of meters after such an initial loss would require substantial water delivery during the late veneer [e.g. *Chou* 1978, *Drake & Richter* 2002, *Frank* et al. 2012].

One surprising observation is that Venus, while being smaller than the Earth, has **more than** twice as much nitrogen in its atmosphere as Earth: $1.1x10^{19}$ versus $3.9x10^{18}$ kg. These numbers were calculated from http://nssdc.gsfc.nasa.gov/planetary/factsheet/venusfact.html and The Cambridge Handbook of Earth Science Data. It is estimated that Venus may have similar abundant quantities of nitrogen beneath its surface [*Lecuyer et al.* 2000]. At the same time Venus has far more $CO_2$ in its atmosphere than Earth contributing to its high surface temperatures by being a potent greenhouse gas. Venus's atmosphere contains $1.2x10^{20}$ kg of carbon, while all superficial Earth reservoirs (the largest being carbonate rock) combined contain $5.4x10^{19}$ kg carbon [*Lecuyer et al.* 2000].

The history of Venus's orbital state is more uncertain. Currently Venus has a 116 Earth-day-length solar day. It orbits the Sun every 225 days and has a 'retrograde' sidereal rotation period of 243 days. It had been thought that atmospheric tides in the thick Venusian atmosphere eventually led to the present rotation rate of Venus [*Dobrovolskis & Ingersoll* 1980], and that this condition was not likely to be primordial. However, it has recently been shown that even a 1 bar atmosphere is sufficient to create the same tidal effect [*Leconte et al.* 2015]. Other work has shown that approaching a tidally locked or nearly tidally locked state like that of modern Venus may be a natural consequence of planetary tidal-bulge/sun interaction [*Barnes* 2016]. New work by Barnes et al. [2016] has shown that if Venus started with its current obliquity near 180° it is likely to have remained so throughout its existence. How it would have obtained its initial ~180° obliquity is still debated.

In this paper we use a 3-dimensional General Circulation Model (GCM) to explore scenarios under which an ancient Venus with shallow oceans and an Earth-like atmosphere may have been habitable and to estimate the potential duration of such a habitable phase.

**Methods**

Several hypothetical Venus climates were simulated (see Table 1) via the Goddard Institute for Space Studies ROCKE−3D (Resolving Orbital and Climate Keys of Earth and Extraterrestrial Environments with Dynamics) GCM. ROCKE-3D is derived from the parent Earth climate GCM ModelE2-R [*Schmidt et al.* 2014]. The simulations were run on a Cartesian grid point system at 4°×5° latitude-longitude resolution, and 20 atmospheric layers with a top at 0.1 hPa. The atmosphere is coupled to a 13 layer fully dynamic ocean [*Russell et al.* 1995]. Land albedo is initially set to 0.2 following *Yang et al.* [2014]. There is no land ice at model start, but snow is allowed to accumulate. Given the fact that Venus shows substantial $N_2$ in its atmosphere today and has few modern day sources or sinks (unlike Earth), we assume that an ancient Venus could have had a ~1 bar $N_2$ atmosphere (1012.6 mb) in its early history. A modern Earth amount of $CO_2$ and $CH_4$ is also included (400ppm, 1ppm), given otherwise poor constraints on these gas concentrations. A variety of ancient solar spectra were generated (Table 1, column 3) with a model by *Claire et al.* [2012] using a modern Sun reference spectrum: 0.5–2397.5 nm [*Thuillier et al.* 2004], and 2397.5-99999.5 nm [*Lean* 2000]. The faintest spectrum at 2.9 Gya is ~77% of the present day luminosity of the Sun.

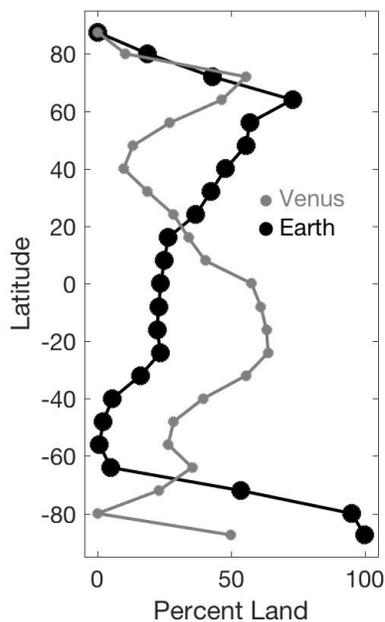

**Figure 1.** Land fraction as a function of latitude for modern Earth (Sim C) versus our paleo-Venus reconstruction (Sims A, B, and D). The paleo-Venus world has a global land/ocean fraction of ~40/60% while modern Earth's is ~30/70%.

Venus was resurfaced by volcanic activity hundreds of millions of years ago [e.g. *McKinnon et al.* 1997, *Kreslavsky & Ivanov* 2015], so its topography before that time is unknown. As an estimate with some observational basis, we use modern topographic data from the Venus Magellan mission via the PDS (Planetary Data System archive http://pds-geosciences.wustl.edu/mgn/mgn-v-rss-5-gravity-l2-v1/mg_5201) and fill the resurfaced lowlands with water. The ocean level was set to the mean radius of Venus from the same PDS data. This provides an ocean of 310 m depth if spread evenly across the surface, fortuitously within the range estimated in *Donahue & Russell* [1997]. The ocean volume is $1.4 \times 10^{17} m^3$, an order of magnitude smaller than that of modern Earth at $1.3 \times 10^{18} m^3$. If an equivalent depth Earth ocean were to exist on modern Venus, less than 10% of the surface would still be land [*Taylor* 2016]. While Earth's ocean covers ~70% of its surface area, this paleo-Venus ocean covers only 60%, with a latitudinal distribution of land distinct from that of modern Earth (Fig. 1). The impact on atmospheric dynamics and thermodynamics of the distribution of land by latitude and land/ocean fraction has seldom been explored [*Abe et al.* 2011, *Charnay et al.* 2013], and so we repeat our baseline simulation (Sim A) with identical forcing but with Earth's topography (Sim C). All simulations use Venus' radius, gravity, modern obliquity, and modern eccentricity.

| Simulation | Topography | Spectrum/Flux/S0X | Rotation Period | T (min) | T (max) | T (avg) | Time (avg) |
|---|---|---|---|---|---|---|---|
| A | Venus | 2.900 Gya/2001/1.46 | Modern Venus | -22˚C | 36˚C | 11˚C | 1/6 day |
| B | Venus | 0.715 Gya/2357/1.70 | Modern Venus | -17˚C | 35˚C | 15˚C | 1/6 day |
| C | Earth | 2.900 Gya/2001/1.46 | Modern Venus | -13˚C | 46˚C | 23˚C | 1/6 day |
| D | Venus | 2.900 Gya/2001/1.46 | 16 x Earth | 27˚C | 84˚C | 56˚C | 1/8 day |

**Table 1.** Paleo-Venus experimental setups and results. The Spectrum/Flux/S0X is the solar spectrum used (Gya=$10^9$ years ago), the incident solar flux the world receives in Watts/$m^2$, and the ratio, S0X, of the solar irradiance the world receives compared to that of present day Earth. T indicates the min and max surface temperature values from individual grid cells in the time averages shown in Figure 2. T (avg) is the globally averaged surface air temperature. Time (avg) is the time period in solar days for each world over which the global temperatures are averaged as indicated in the temperature columns.

To test the range of ages and rotations over which the climate of paleo-Venus might have been stable, paleo-Venus was also simulated using a solar spectrum estimated for 0.715Gya (Sim B), equivalent to ~94% of modern day solar luminosity (Table 1, Fig. 2b) and a solar flux at Venus's orbit ~70% higher than that for modern Earth. The value of 0.715Gya was chosen to represent the time slice around the time of the most recent resurfacing event on Venus. To test the possible effects of rotation, Sim D (Fig. 2d) uses the same 2.9Gya solar spectrum as in Sim A, but changes the rotation period from the modern Venus value to 16 times that of modern Earth's sidereal day length.

**Results**

Sim A gives a global mean surface air temperature of 11˚C (Table 1, Fig. 2a) and maximum of 36˚C in a single grid cell despite an incident solar flux at Venus's orbit ~40% higher than that received by modern Earth. The range of surface temperatures in Sim A is narrower than that of modern Earth (Fig. 3), primarily because dynamical heat transports more efficiently reduce horizontal temperature gradients on slowly rotating planets and secondarily because Earth's lowest temperatures occur in its polar continental regions that Sim A lacks. Previous 1-D work by *Grinspoon & Bullock* [2007] has shown that an $N_2$ atmosphere with 6 mb of $H_2O$ (at the surface) can generate a surface air temperature of 27˚C if 100% cloud cover is assumed. Their 100% cloud cover assumption is nicely reflected in that of Sim A where the dayside of the planet is almost completely overcast (Fig. 4a).

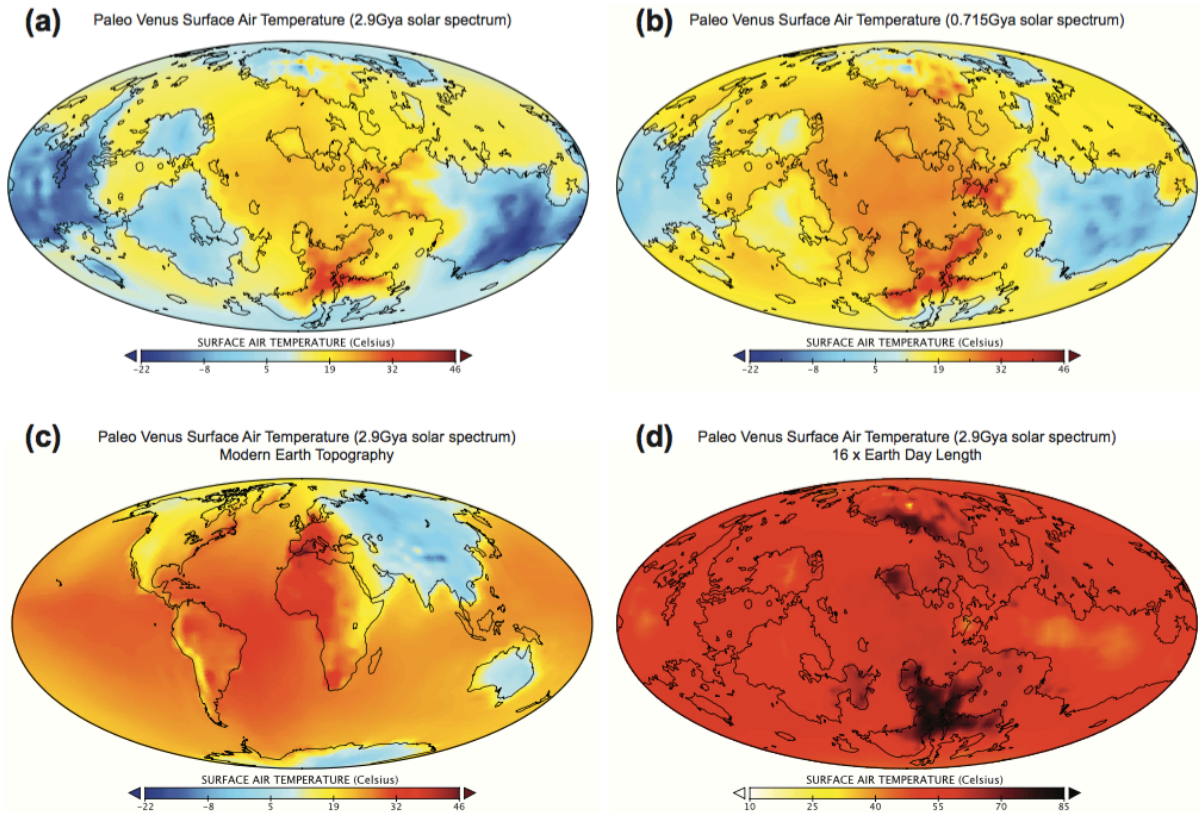

**Figure 2.** Top left (a) is a time average (over $1/6^{\text{th}}$ of a Venusian solar day) of the surface air temperature (SAT) for the paleo-Venus world with modern day rotation rate and a solar spectrum estimated for 2.9Gya (Sim A). Top right (b) is the SAT for the same world in 2a, but using a 0.715Gya solar spectrum (Sim B). Bottom left (c) is the same as (a) but using a modern day Earth topography (Sim C). Bottom right (d) uses a 2.9Gya solar spectrum but contrary to the other 3 simulations has a sidereal day 16 times that of modern Earth (Sim D). The snapshot shown in (d) is an average over $1/8^{\text{th}}$ of a solar day on this world, again contrary to the other 3 simulations. Note that the color scale in 2(d) has changed from that of 2(a,b,c). The yellow color in 2(d) has a similar value to the yellow in 2(a,b,c) (around 17-19˚C). Sim D reaches the limit of the accuracy of the current ROCKE-3D radiation tables, which use a simple linear extrapolation above 77˚C.

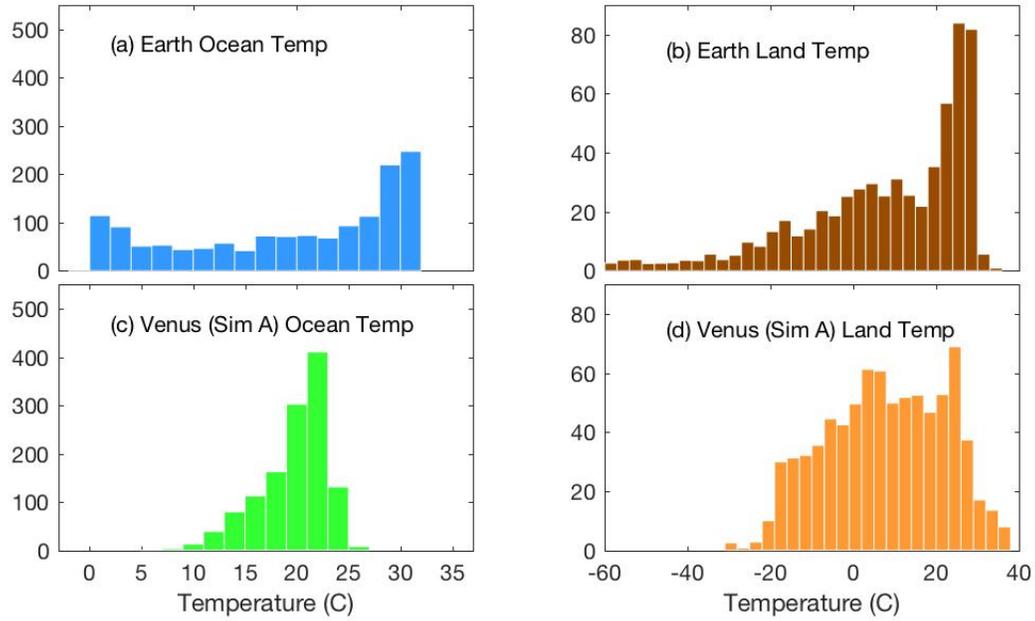

**Figure 3.** Left panels are equal area histograms of ocean surface temperatures for (a) modern Earth [*Schmidt et al.*, 2014] and (c) Sim A. Right panels contain land temperatures for (b) modern Earth and (d) Sim A.

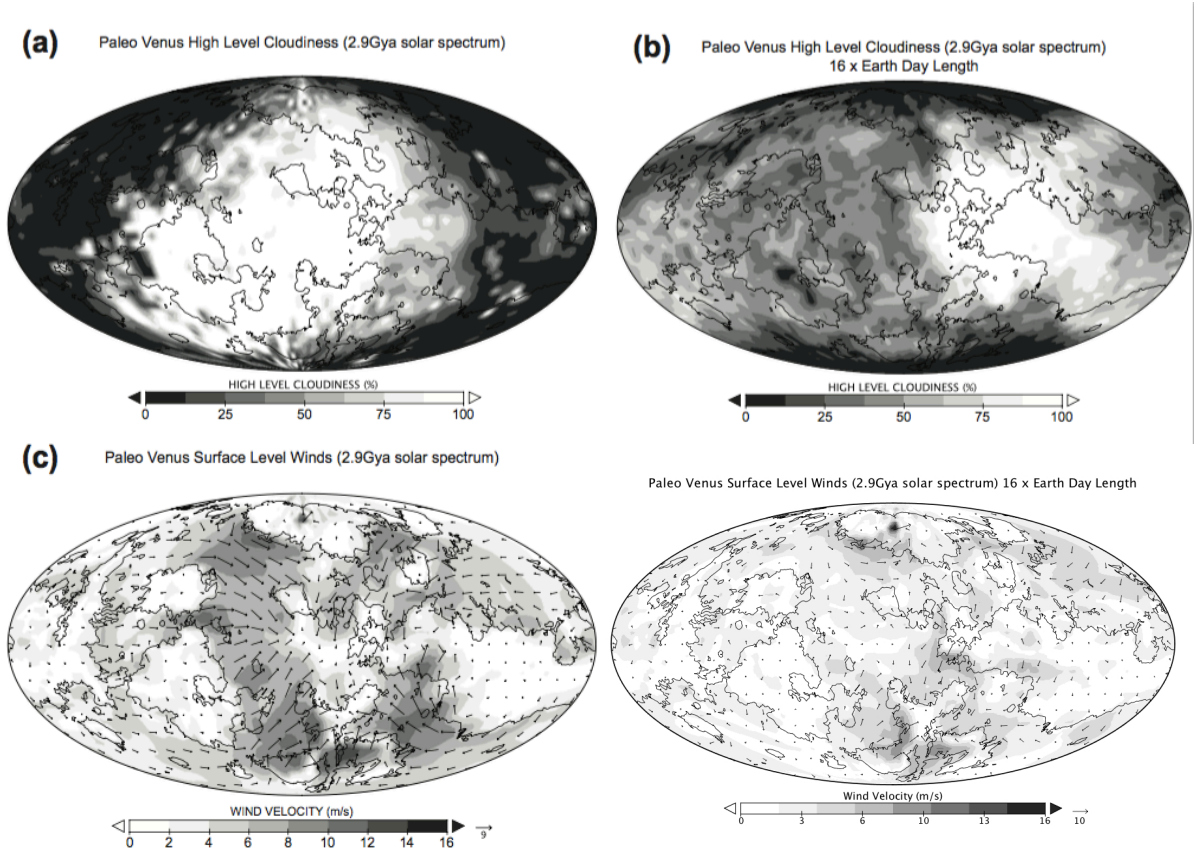

**Figure 4.** High level clouds (white) for Sim A (a) and D (b), and surface level winds for Sim A (c) and D (d). Shading in 4(c,d) is related to wind strength.

Comparing Sims A and B, where the solar flux increases from 77% to 94% of modern, one sees that the global mean surface air temperature increases only 4°C. However, the effect of rotation on temperature is much stronger when the world's rotation period is decreased to 16 times the present day sidereal rotation rate of modern Earth, but the solar spectrum remains from 2.9Gya (Sim D, Fig. 2d). The faster rotation rate gives Sim D a mean surface air temperature 45°C warmer than that of Sim A. This should be unsurprising considering the similar results of *Yang et al.* [2014, their Fig. 2c], who also find a pronounced surface air temperature difference between their 16 and 256 sidereal day length worlds.

The high-level cloudiness on the dayside of Sim A (Fig. 4a) has values as high as 100% for a given grid cell. This is the result of this world's slow rotation which generates a strong circulation with rising motion (implied in the surface wind plot of Fig. 4c) and accompanying high thick clouds on the dayside that reflect a substantial fraction of the incident sunlight, as also shown in experiments by *Yang et al.* [2014]. These highly reflective thick areally extensive clouds are formed from smaller scale parameterized convective clouds that carry water vapor and condensed water to high altitudes in buoyant updrafts. When these updrafts reach the stable middle/upper troposphere they lose buoyancy and the flow diverges and spreads, injecting water vapor and cloud particles that form a reflective shield of cirrus anvil clouds. This is similar to much of the cloudiness seen in the Intertropical Convergence Zone on Earth, except that the convection in our paleo-Venus worlds sits more or less in place for many days at a time since the rotation is so slow and can make a large area overcast for several months.

This circulation pattern is peculiar to the assumed Earth-like thickness of the paleo-Venus atmosphere, which allows it to cool significantly on the night-side by emitting thermal radiation [*Way et al.* 2015] generating a strong day-night circulation (Fig. 4c). The latter may be contrasted with the wind pattern in Figure 4d for the 16 d rotation period case (Sim D) where the surface level winds do not show a strong day-night flow. Modern Venus, which has a much thicker $CO_2$ atmosphere that takes centuries to cool radiatively, has only a weak day-night circulation and is dominated instead by a broad Hadley cell that stretches from the equator to high latitudes [*Rossow et al.* 1990].

The simulation with Earth topography (Sim C) produces habitable surface temperatures but is 12°C warmer than the otherwise equivalent Sim A with Venus topography (Table 1, Fig. 2). This result can be explained by the finding by *Abe et al.* [2011] that "dry" planets with limited water reservoirs are less sensitive to increases in solar flux than all-water "aquaplanets." In particular, the large land fraction in the tropics in the Venus topography case relative to that for Earth topography (Fig. 1) appears to limit surface evaporation (which is largest over the tropical oceans) and thus the buildup of atmospheric water vapor and its attendant greenhouse effect at low latitudes.

From Figure 2 it is clear that the night side of the slowly rotating Sims A, B and C are cold enough for snow to accumulate. In most places this snow amounts to a few centimeters and completely melts by the time the Sun is overhead. We find no trend in this behavior over the length of the simulations that would lead to a large cold trap. However, in Sims A and B the two highest altitude grid cells (over 5000m) in Ishtar Terra (high northern latitude) have a permanent snow cover of about 5 meters in depth which appears constant in time.

One might also believe that the land would get quite dry from repeated heating, yet the water in the first ground layer (10cm) has a relatively zonal distribution and does not appear to be drying out due to the very long periods of solar heating in the slowly rotating cases. In fact the ground appears to get wet during the day and stays mostly wet throughout the night.

The mean precipitation rate is at least 8mm/day in the subsolar region, with 10s of mm/day in some grid boxes for Sim A. This is as strong or stronger than anything seen on Earth, The potential for weathering and erosion on such a planet would exist, but only within ~20° of the equator (unlike on Earth). Hence a future mission to Venus should explore Aphrodite Terra and Beta Regio, rather than Ishtar Terra, to search for such erosion. Visual signs of such water-related erosion and transport of sediment (e.g., stream down-cutting through land forms, gravelly/braided stream deposits, mudslides, alluvial fans or deltas) would indicate physical erosion mainly. Both of these are potentially detectable via remote sensing, just as they have been for Mars. Indications that chemical weathering had taken place would include the presence of certain clays, depending on the composition of the parent rock. However, rates of chemical weathering are difficult to constrain, given Venus' extreme long-term climate variability.

If these hypothetical worlds were in or were approaching a moist greenhouse state then it could be possible to conclude something quantitative about the likelihood of water loss over the history of the planet. *Kasting* [1988] showed that this state would be attainable if the mixing ratio of $H_2O$ in the upper atmosphere reached ~0.1%. The paleo-Venus Sim A world is well below that mixing ratio, and its stratosphere (100mb level and higher) is as dry as modern Earth, nearly 2 orders of magnitude below the moist greenhouse limit. Sim B is one order of magnitude below, and Sim C is half-way between Sims A and B, while Sim D is 1 order of magnitude over the limit.

**Discussion**

Several caveats need to be discussed in the context of the ancient climates of Venus, Earth and Mars. Early Earth and Mars would have been very cold worlds circa 2.9 Gya without substantial greenhouse gases or other means to absorb and re-radiate thermal radiation [*Goldblatt & Zahnle* 2010, *Charnay et al.* 2013, *Ramirez et al.* 2014]. Another item seldom mentioned in studies of ancient Mars and the possibility of life is the issue of climate stability: a mechanism to keep the sinks and sources of greenhouse gases in balance for long enough periods of time to allow life to begin. The same holds true for Venus. Venus has substantial sources of $N_2$ and $CO_2$ as measured in its atmosphere today, but no substantial sinks have been measured thus far. Studies since the time of the Magellan mission have shown that subductive style plate tectonics do not presently exist on Venus to give it a substantial sink like that on Earth, even if there may be other types of tectonic activity allowing sinks of unquantified magnitude [*Elkins-Tanton et al.* 2007]. So while this work demonstrates that it may have been possible to have a moderate climate on Venus on short time scales compared with Earth [*Sleep & Zahnle* 2001, *Bertaux et al.* 2007], we must wait until future data help us better constrain that planet's ancient geologic history to understand if it was capable of long-term stable habitable climates. Other anticipated spacecraft measurements will reduce current large uncertainties in the geologic and atmospheric history of Venus. High resolution radar and global near-infrared mapping will help clarify both ancient and recent levels of geologic activity, informing comparisons with terrestrial tectonic style and history [*Smrekar et al.* 2016]. *In situ* measurements of rare gases and their isotopes, currently highly uncertain, as well as precision measurements of D/H and other stable isotope ratios, will greatly refine current constraints on the atmospheric and climatic history of this nearby Earth-sized planet [*Glaze et al.* 2016].

Since Venus is similar to Earth in size and adjacent to it in our solar system, it is of particular interest for the characterization of seemingly similar exoplanets that could have followed different evolutionary paths [*Kane et al.* 2014]. It is not known with certainty whether Venus ever had an ocean. Venus could have formed with less water than Earth [*Raymond et al.* 2006], or with considerable water that escaped during an extended early runaway greenhouse period during formation [*Hamano et al.* 2013, *Luger & Barnes* 2015]. Yet these results and others [*Yang et al.* 2014] suggest that for warm rocky planets that retain significant water after formation, an extended period of habitability is possible well inside the traditionally defined inner edge of the habitable zone for cloud-free atmospheres [*Kopparapu et al.* 2013] if they rotate slowly enough. Furthermore, while the possibility of surface liquid water defines the traditional habitable zone, our results suggest that a planet with a modest amount of surface liquid water is more conducive to habitability over a wide range of stellar fluxes than a planet largely or completely covered by water. The inner edge should therefore be considered a transition region in which the probability of habitability gradually decreases inward rather than a strict boundary separating completely different regimes [*Barnes et al.* 2015].

## Acknowledgements

This research was supported by the NASA Astrobiology Program through the Nexus for Exoplanet System Science (NExSS) research coordination network sponsored by NASA's Science Mission Directorate. This work was also supported by NASA Goddard Space Flight Center ROCKE-3D Science Task Group funding. Resources supporting this work were provided by the NASA High-End Computing (HEC) Program through the NASA Center for Climate Simulation (NCCS) at Goddard Space Flight Center. This research has made use of NASA's Astrophysics Data System Bibliographic Services. Thanks to Jeffrey A. Jonas, Kostas Tsigaridis and David S. Amundsen for their assistance in this work.